\documentstyle[epsfig]{aipproc}

\begin{document}
\title{QCD analysis of $F_2^{\gamma}(x,Q^2)$: an unconventional view
\footnote{Talk given at PHOTON2000 Symposium, Ambleside, UK, August 
2000}}

\author{Ji\v{r}\'{\i} Ch\'{y}la$^*$}
\address{$^*$
Center for Particle Physics, Institute of Physics
of the Academy of Sciences\\
18221 Na Slovance 2, Prague 8, Czech Republic}
%
%Institute of Physics, Academy of Sciences, Na Slovance 2, Prague 8,
%Czech Republic}
%\lefthead{LEFT head}
%\righthead{RIGHT head}
\maketitle
\begin{abstract}
Elements of the alternative approach to hard collisions of photons,
proposed recently by the author, are reviewed, with particular 
attention to QCD analysis of $F^{\gamma}_2$. This approach is based 
on clear separation of genuine QCD effects from those of pure QED 
origin and does not rely on the assumption that parton distribution 
functions of the photon behave as $\alpha/\alpha_s$. It differs 
significantly from the conventional one, as illustrated on the example 
of charm contribution to $F_{2}^{\gamma}$, recently measured at LEP.
\end{abstract}

\section*{Introduction}
\label{intro}
\noindent
In \cite{jfactor} I have proposed an alternative approach to QCD 
analysis of $F^{\gamma}_2(x,Q^2)$, and by implication to any hard process 
involving initial photon, which differs substantially from the 
conventional one. It builds in part on arguments advocated for a long 
time by the authors of \cite{FKP} and agrees with the approach to 
calculations of direct photon production at HERA pursued in \cite{Maria}.
This alternative approach is based on two ingredients:
\begin{itemize}
\item Clear and systematic separation of genuine QCD effects from
those of pure QED origin, which leads to unambiguous and universal
definition of the concepts ``leading'' and ``next--to--leading'' order
of QCD.
\item Acknowledgement of the fact that parton distribution functions 
(PDF) of the photon are proportional to $\alpha$ and not, as assumed in 
the conventional approach, to $\alpha/\alpha_s$.
\end{itemize}
The general expression for $F_2^{\gamma}(x,Q^2)$ has the following 
structure
\begin{eqnarray}
\frac{1}{x}F_2^{\gamma}(x,Q^2)&=& q_{\mathrm{NS}}(M)\otimes
C_q(Q/M)+\frac{\alpha}{2\pi}\delta_{\mathrm{NS}}C_{\gamma}+
\langle e^2\rangle \Sigma(M)\otimes
C_q(Q/M)+\label{ok}\nonumber \\
& & \frac{\alpha}{2\pi} \langle
e^2\rangle\delta_{\Sigma}C_{\gamma}+\langle
e^2\rangle G(M)\otimes C_G(Q/M)
\label{F2}
\end{eqnarray}
where quark nonsinglet and singlet and gluon distribution functions
\footnote{For their definitions as well as other details of the 
alternative approach, see \cite{jfactor}.}
satisfy the evolution equations
\begin{eqnarray}
\frac{{\mathrm d}\Sigma(x,M)}{{\mathrm d}\ln M^2}& =&
\delta_{\Sigma}k_q+P_{qq}\otimes \Sigma+ P_{qG}\otimes G,
\label{Sigmaevolution}
\\ \frac{{\mathrm d}G(x,M)}{{\mathrm d}\ln M^2} & =& k_G+
P_{Gq}\otimes \Sigma+ P_{GG}\otimes G, \label{Gevolution}\\
\frac{{\mathrm d}q_{\mathrm {NS}}(x,M)}{{\mathrm d}\ln M^2}& =&
\delta_{\mathrm {NS}} k_q+P_{\mathrm {NS}}\otimes q_{\mathrm{NS}},
\label{NSevolution}
\end{eqnarray}
where $\delta_{\mathrm{NS}}=6n_f\left(\langle e^4\rangle-\langle
e^2\rangle ^2\right),\delta_{\Sigma}=6n_f\langle e^2\rangle$.
The splitting and coefficient functions $k_i,C_j$ can be expanded 
in powers of $\alpha_s$ ($a(M)\equiv \alpha_s(M)/2\pi$) as
\begin{eqnarray}
k_q(x,M) &= & \frac{\alpha}{2\pi}\left[{  k^{(0)}_q(x)}+
a(M)k_q^{(1)}(x)+a^2(M)k^{(2)}_q(x)+\cdots\right],
\label{splitquark} \\
k_G(x,M) & = &\frac{\alpha}{2\pi}\left[
a(M)k_G^{(1)}(x)+a^2(M)k^{(2)}_G(x)+\cdots\;\right],
\label{splitgluon}\\
P_{ij}(x,M) & = & a(M)P^{(0)}_{ij}(x)+ 
a^2(M) P_{ij}^{(1)}(x)+\cdots,
\label{splitpij}
\end{eqnarray} 
\begin{eqnarray}
C_q(x,Q/M) & = & {  \delta(1-x)}+a(M)C^{(1)}_q(x, Q/M)+
a^2(M)C^{(2)}_q(x, Q/M)+\cdots,
\label{cq} \\
C_G(x,Q/M) & = & a(M)C^{(1)}_G(x,Q/M)+a^2(M)C^{(2)}_G(x,Q/M)+\cdots,
\label{cG} \\
C_{\gamma}(x,Q/M) & = & {  C_{\gamma}^{(0)}(x,Q/M)}+
a(M)C_{\gamma}^{(1)}(x,Q/M)+
a^2(M)C_{\gamma}^{(2)}(x,Q/M)\cdots,
\label{cg}
\end{eqnarray}
where the lowest order coefficient functions $k_q^{(0)}$ and 
$C_{\gamma}^{(0)}$ 
\begin{eqnarray}
 k_q^{(0)}(x) & = & (x^2+(1-x)^2), \label{k0}\\
C_{\gamma}^{(0)}(x,Q/M)& = &\left(x^2+(1-x)^2\right)
\left[\ln\frac{M^2}{Q^2}+\ln\frac{1-x}{x}\right]+8x(1-x)-1
\label{c0}
\end{eqnarray}
are unique
\footnote{Throughout this paper we restrict ourselves to the case
of real target photon, although the basic conclusions hold for the
virtual photon as well. For the latter, however, the expression 
for $C_{\gamma}^{(0)}$ differs slightly from (\ref{c0}).}
 and due entirely to QED coupling of the initial photon to the primary 
$q\overline{q}$ pair. It is their presence what distinguishes hard 
collisions of photons from those of hadrons. In \cite{jfactor} I have 
discussed conceptual as well as numerical differences between the 
results obtained within the conventional and alternative approaches 
for the pointlike part of $F_2^{\gamma}$ in the nonsinglet channel and
under the simplifying assumption $\beta_1=0$.
In this talk I will concentrate on proper definitions of the concepts 
``leading order'' (LO) and ``next--to--leading order'' (NLO) of QCD
and on the discussion of their implications for charm contribution 
$F_{2,c}^{\gamma}(x,Q^2)$ to photon structure function, recently 
measured at LEP \cite{OPAL,Richard}.

\section*{Defining leading and next-to-leading orders of QCD}
Although the definition of the concepts ``LO'' and ``NLO'' in hard 
collisions of photons is a matter of convention, it is preferable to 
define them in a way that retains the same basic content these concepts
have in collisions of leptons and hadrons. Let me recall, for instance,
their meaning in the case of the familiar ratio
\begin{equation}
R_{\mathrm{e}^+\mathrm{e}^-}(Q)\equiv
\frac{\sigma(\mathrm{e}^+\mathrm{e}^-\rightarrow
\mathrm{hadrons})}{\sigma(\mathrm{e}^+\mathrm{e}^-\rightarrow
\mu^+\mu^-)}
=\left(3\sum_{i=1}^{n_f}e_i^2\right)\left(1+r(Q)\right).
\label{Rlarge}
\end{equation}
The prefactor $R_{\mathrm{QED}}\equiv 3\sum_{i=1}^{n_f}e_i^2$
multiplied by unity in the brackets of (\ref{Rlarge}) comes purely from 
QED, whereas genuine QCD effects are contained in $r(Q)$, which is given 
as expansion in powers of $\alpha_s$ as
\begin{equation}
r(Q)=\frac{\alpha_s(M)}{\pi}\left[1+\frac{\alpha_s(M)}
{2\pi}r_1(Q/M)+\cdots\right].
\label{rsmall}
\end{equation}
For the quantity (\ref{Rlarge}) it is a generally accepted practice 
to include $R_{\mathrm{QED}}$ in theoretical expressions but disregard it 
when defining the ``LO'' and ``NLO'' of QCD. 
For instance, the NLO approximation of (\ref{Rlarge}) implies retaing
{\em first two} terms in (\ref{rsmall}), i.e. the {\em first three} terms 
in (\ref{Rlarge}). Let me emphasize that only if at least first two 
nontrivial powers of $\alpha_s$ in (\ref{Rlarge}) are taken into account 
can this trunctated expansion be associated with a well-defined 
renormalization scheme. And it is this association what makes in my view 
the essential feature of the concept ``NLO QCD approximation''. The same 
convention should be adopted for all physical quantities getting 
contributions from pure QED. I think it is preferable to use the 
terminology that avoids  potential confusion which might arise from 
mixing orders of $\alpha_s$ and $\alpha$.
Note that for both $R_{\mathrm{e}^+\mathrm{e}^-}$ and 
$F_{2,c}^{\gamma}(x,Q^2)$, discussed in the next Section, the purely QED 
contributions $R_{\mathrm{QED}}$ and $F_{2,c}^{\gamma,{\mathrm{QED}}}$
are finite and unique and there is thus no reason why they should be 
treated differently as far as the definitions of the ``LO'' and 
``NLO'' of QCD are concerned. 
The implications of the above considerations for $F_2^{\gamma}(x,Q^2)$
are the following: 
\begin{itemize}
\item As shown in \cite{jfactor} the LO QCD expression for $F_2^{\gamma}$
contains in addition to terms included in the conventional LO analysis of 
$F_2^{\gamma}$ (i.e. those proportional to $k_q^{(0)}$ and $P^{(0)}_{ij}$), 
also terms involving $k_q^{(1)},C_q^{(1)},C_{\gamma}^{(0)}$ and 
$C^{(1)}_{\gamma}$. As all these functions are known, there is, however, 
no obstacle to performing such an analysis. As shown in \cite{jfactor} for 
the pointlike part of $F_2^{\gamma}(x,Q^2)$ in the nonsinglet channel, the 
results in these two approaches are numerically significantly different, 
the single most important contribution to this difference coming from 
$C_{\gamma}^{(1)}$. The coefficient function $C_q^{(1)}$ enters 
$F_2^{\gamma}$ already at the LO
\footnote{Contrary to hadronic collisions, where it appears first at the
NLO.}
due to the fact that it does so in the 
convolution with purely QED part of quark distribution function of the 
photon, which has no analogue in hadronic collisions.
\item  
The NLO QCD analysis of $F_2^{\gamma}(x,Q^2)$ requires the knowledge of 
two quantities, $k_q^{(2)}$ and $C_{\gamma}^{(2)}$, that have not yet been 
calculated and thus is at the moment impossible to perform. Note, that in 
the conventional approach $C_{\gamma}^{(0)}$, which has nothing to do with 
QCD, appears only at the ``NLO'', and $C_{\gamma}^{(1)}$, which involves 
evaluation of Feynman diagrams with a single QCD vertex is not used even 
there!
\end{itemize}
\begin{figure}\centering
\epsfig{file=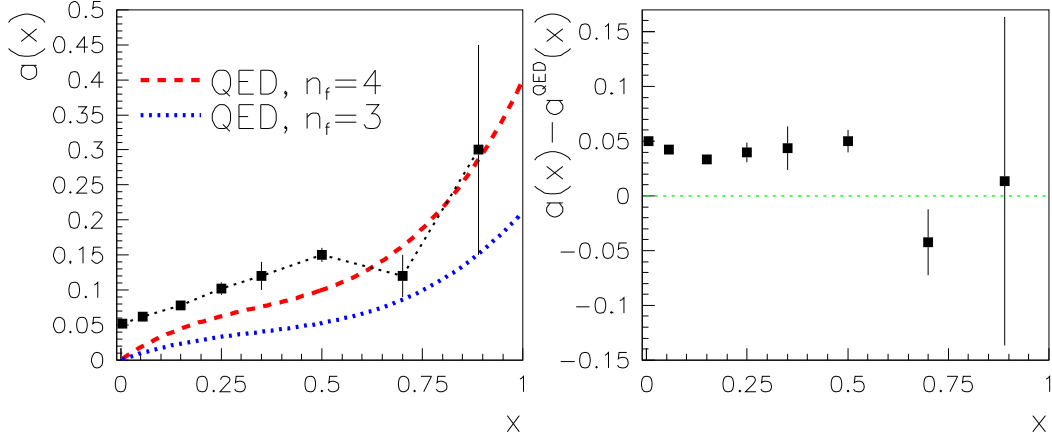,width=\textwidth}
\caption{The derivative (\ref{a}) taken from from the fit to LEP data
(solid squares) compared to pure QED formula
$a^{\mathrm{QED}}(x)=3\sum_{i=1}^{n_f}e_i^4x(1+(1-x)^2)$ for $n_f=3,4$.
Both data and curves in units of $\alpha$.}
\label{slopes}
\end{figure}
For clear and unambiguous definition of the terms ``LO'' (``NLO'') it 
is thus vital to agree on the basic criterion, namely that they refer to 
perturbative expansions of physical quantities retaining the first 
(first two) {\em nontrivial} powers of $\alpha_s$. The purely QED 
contribution to $F_2^{\gamma}(x,Q^2)$ is irelevant from this point of
view, but may be retained for comparison with experiment as it actually 
dominates scaling violations of $F_{2}^{\gamma}(x,Q^2)$ in most of 
accessible range of $x$. To identify genuine QCD effects one has to look 
for subtler effects than the dominant $\ln Q^2$ rise, like, for instance, 
the $x$-dependence of the slope \begin{equation}
a(x)\equiv
\frac{{\mathrm{d}}F_{2}^{\gamma}(x,Q^2)}
{{\mathrm{d}}\ln Q^2}.
\label{a}
\end{equation}
Compared to $F_2^{\mathrm{p}}(x,Q^2)$, for which scaling violations are
due entirely to QCD effects, the nonzero slope (\ref{a})
is by itself no sign of QCD effects, as these are given by the difference 
$\Delta(x)\equiv a(x)-a^{\mathrm{QED}}(x)$. Fig. \ref{slopes}, based on
numbers taken from \cite{Richard2}, shows that 
for $x\gtrsim 0.5$ the precision of currently available data is 
insufficient to identify 
genuine QCD effects, although some indication of the turnover to negative 
$\Delta(x)$, expected theoretically, is visible there.

\section*{Charm contribution to 
$F^{\gamma}_2(\protect{{\small {x}}},Q^2)$}
\begin{figure}[t]\centering
\epsfig{file=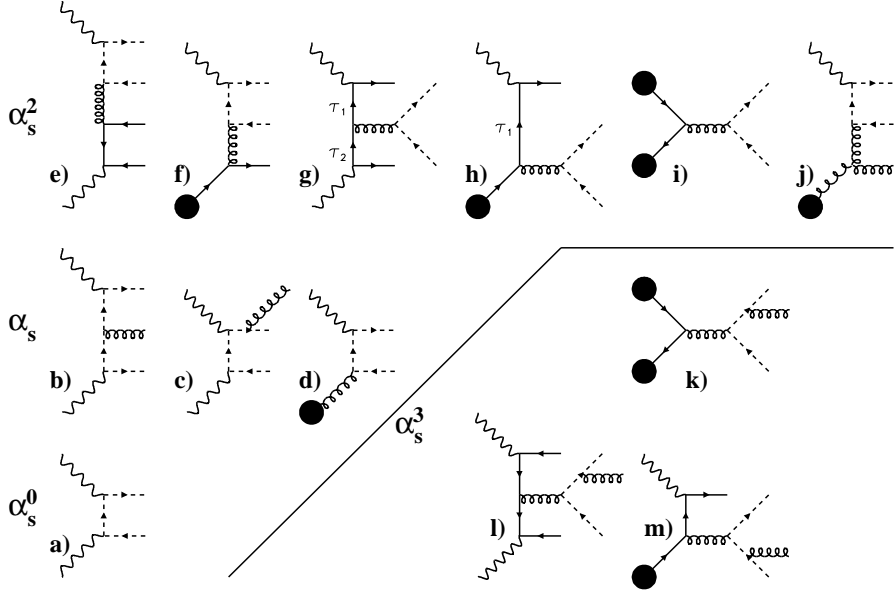,width=12cm}
\caption{Examples of diagrams up to order $\alpha_s^3$ contributing to heavy
quark production in photon-photon collisions. The upper (lower) wavy 
lines corresponds to the probing (target) photon. Light quarks are denoted 
by solid, heavy quarks by dashed lines. For $F_{2,c}^{\gamma}$ the diagrams 
with blobs in the upper vertices are absent as only the target photon may
be ``resolved''. The initial state singularities of the
direct and single resolved photon contributions coming from the vertex
$\gamma\rightarrow q\overline{q}$ are understood to be subtracted and
put into PDF of the photon, described by solid blobs. As a results,
the (subtracted) direct and resolved photon diagrams, as well as
the associated PDF acquire dependence on the factorization scale $M$.}
\label{ggnnlo}
\end{figure}
The QCD analysis of $F_{2,c}^{\gamma}(x,Q^2)$ in the region $x\gtrsim 0.1$ 
provides possibly the simplest illustration of the differences between 
the conventional and alternative approches to hard collisions of photons
\cite{plb}, since $F_{2,c}^{\gamma}$ is dominated in this region by the 
direct photon contribution, $F_{2,c}^{\mathrm{dir}}$, which does not 
involve any PDF. It can be written as 
\begin{equation}
F^{\mathrm{dir}}_{2,c}(x,Q^2)=
F^{\mathrm{QED}}_{2,c}(x,Q^2)+\alpha_s(Q)F^{(1)}_{2,c}(x,Q^2)+
\alpha_s^2(Q)F^{(2)}_{2,c}(x,Q^2)+\cdots,
\label{dir} 
\end{equation}
where the coefficients $F_{2,c}^{(k)},k\ge 1$ are calculable in 
perturbative QCD and the lowest order term 
$F_{2,c}^{(0)}\equiv F_{2,c}^{\mathrm{QED}}$ comes from pure QED diagram in 
Fig. \ref{ggnnlo}a.
In the conventional approach the ``NLO'' approximation of the direct photon 
contribution is defined \cite{Laenen} by taking the first two terms in 
(\ref{dir}) including the purely QED contribution $F^{\mathrm{QED}}_{2,c}$. 
However, as argued in the preceding Section, this definition does not have 
the basic attribute of genuine NLO QCD approximation. The inclusion
of direct photon contributions of the order $\alpha^2\alpha_s^2$, 
coming from diagrams like those in Fig. \ref{ggnnlo}e,g is vital not only 
for establishing the genuine NLO character of the direct photon contribution
itself, but also for ensuring \cite{plb} factorization scale invariance 
of the full expression for $F_{2,c}^{\gamma}$. The latter involves adding 
the resolved photon contributions up to the order $\alpha^2\alpha_s^2$, 
coming from diagrams like those in Fig. \ref{ggnnlo}f,h,j.

\section*{Summary and conclusions}
The alternative approach to QCD analysis of $F_2^{\gamma}$, proposed 
recently by the author, differs substantially and in a number of aspects
from the conventional one. It satisfies factorization scale invariance in 
a way that does not rely on physically untenable assumption that quark
distribution functions of the photon behave as ${\cal O}(\alpha/\alpha_s)$.
The simplest implications of this difference are illustrated on the
case of charm contribution to $F_2^{\gamma}(x,Q^2)$, recently measured at
LEP.
  
To be useful phenomenologically the proposed approach needs to be further 
elaborated by extending it to the singlet sector and merging it with the 
hadronic contributions. Work on this program is in progress. The NLO QCD 
analysis in the proposed approach requires evaluation of several so far 
unknown quantities and is thus currently impossible to perform. In view 
of the quality and number of experimental data on $F_2^{\gamma}$, this is 
at the moment no serious drawback and a complete LO QCD analysis seems 
sufficient for phenomenological purposes.

\vspace*{0.5cm}

Work supported by the Ministry of Education of the Czech Republic
under the project LN00A006.

\end{document}